\documentclass[prl,aps,twocolumn,superscriptaddress]{revtex4}

\usepackage{graphicx}
\usepackage{pstricks}
\usepackage{enumerate}
\usepackage{subfig}
\usepackage{amssymb,amsmath,latexsym}
\usepackage{amsthm}
\usepackage{hyperref}
\usepackage{multirow}
\usepackage{dsfont}
\usepackage{pifont}
\usepackage{bm}
\usepackage{textcomp}
\usepackage{microtype}
\usepackage{url}
\usepackage{units}

\newcommand{\Prob}{\operatorname{Pr}}

\newcommand{\be}{\begin{equation}}
\newcommand{\ee}{\end{equation}}

\newcommand{\setA}{\mathcal{A}}
\newcommand{\setB}{\mathcal{B}}
\newcommand{\setX}{\mathcal{X}}
\newcommand{\setY}{\mathcal{Y}}

\def\01{\{0,1\}}
\theoremstyle{plain}

\theoremstyle{definition}

\theoremstyle{remark}

\begin{document}

\title{Computability limits non-local correlations}
\author{Tanvirul Islam}
\email{tanvirul@nus.edu.sg}
\affiliation{Centre for Quantum Technologies, National University of Singapore, 3 Science Drive 2, 117543 Singapore}
\author{Stephanie Wehner}
\email{wehner@nus.edu.sg}
\affiliation{Centre for Quantum Technologies, National University of Singapore, 3 Science Drive 2, 117543 Singapore}

\begin{abstract}
	If the no-signalling principle was the only limit to the strength of non-local correlations, we would expect that any form of no-signalling correlation can indeed
	 be realized. That is, there exists a state and measurements that remote parties can implement to obtain any such correlation. Here, we show that in any theory in which some functions cannot be computed,
	 there must be further limits to non-local correlations than the no-signalling principle alone.
We proceed to argue that even in a theory such as quantum mechanics in which non-local correlations are already weaker, the question of computability imposes such limits.

\end{abstract}
\maketitle

Bell's seminal work~\cite{bell} on non-local correlations, not only allowed us to distinguish classical from quantum mechanics, but also spurred a multitude of useful applications in quantum information theory (see e.g.~\cite{e91,andreas:zeroErrorCap}). As a result it is important to know which non-local correlations can indeed be physically realized.

Let us now explain more carefully what is meant by non-local correlations. For simplicity, we thereby consider a bipartite system, Alice and Bob. 
Let $\setX$ and $\setY$ denote a set of possible measurements for Alice and Bob respectively, and let $\setA$ and $\setB$ denote corresponding set of outcomes.
Furthermore, let
\begin{align}
	\Pr[a,b|x,y]
\end{align}
denote the probability that Alice and Bob obtain outcomes $a \in \setA$ and $b \in \setB$ when making measurements $x \in \setX$ and $y \in \setY$ respectively.
When it comes to the study of non-local correlations, we are interested in the set of allowed values for $\Pr[a,b|x,y]$. For example, the famous CHSH inequality~\cite{chsh} tells us that for any classical theory
\begin{align}\label{eq:chsh}
	\frac{1}{4} \sum_{x,y \in \01} \sum_{a \in \01} \Pr[a,b = x\cdot y \oplus a|x,y] \leq \frac{3}{4}\ ,
\end{align}
for any possible measurements labeled $\setX = \01$ and $\setY = \01$, and any state shared by Alice and Bob~\footnote{$\oplus$ denotes addition modulo 2.}.
Quantumly, however, there exist a shared state and measurements such that the sum attains the value of $1/2 + 1/(2\sqrt{2}) \approx 0.853$. This is the highest value possible in quantum mechanics~\cite{tsirel:original}. It demonstrates that
in a quantum world, non-local correlations can be strictly stronger than is allowed classically. Much work has gone into determining bounds on quantum
correlations~\cite{tsirel:original,landau:compat,landau:bell,andrew:states,wehner05d,npa1,npa2,dltw}, and investigating the difference between the classical and quantum regime (see e.g.~\cite{jp:violation,jppvw:unbounded}).

Here, we are interested in the following question: even if non-local correlations are consistent with a particular theory, should we expect them to be physical? That is, do we really expect them exist in the sense that Alice and Bob can realize the joint state and measurements needed to obtain them?
To explain this question, let us first consider the simple scenario 
of a world in which the only constraint on non-local correlations is that they must be no-signalling~\cite{popescu:nonlocal,popescu:nonlocal2,popescu:nonlocal3}. 
That is, we have
\be
 \label{eq:nosig}\forall a,x,y,y' \quad\Prob[a|x,y] = \Prob[a|x,y']\ ,
\ee
and similarly for all $x,x'$.  Property \eqref{eq:nosig} tells us that, by observing the outcome $a$, Alice can never tell what Bob's input was and vice-versa.  

An example of such maximally correlated pair of nonlocal systems is the famous PR-Box~\cite{popescu:nonlocal,popescu:nonlocal2,popescu:nonlocal3}, 
which allows the violation of the CHSH inequality~\eqref{eq:chsh} up to the value $1$. When thinking about a PR-box, one can think of the measurements more abstractly and imagine that Alice and Bob each have a terminal 
in which Alice can input a bit $x$ and get as outcome a bit $a$. Similarly, Bob can input a bit $y$ and get bit $b$ as his outcome. The distribution of outcomes $a$ and $b$ given the inputs $x$ and $y$ is:
\be
\label{eq:pr}\Prob[a,b|x,y] = \begin{cases}
				0 & \text{if } a\oplus b\neq x\cdot y,\\
				\frac{1}{2} & \text{otherwise},
			\end{cases}
\ee
That is, the inputs $x,y \in \{0,1\}$ and outputs $a,b\in \{0,1\}$ always satisfy
\be
\label{eq:prnice}a \oplus b = x\cdot y.
\ee
One can see, from equation \eqref{eq:pr}, that the marginal of this distribution satisfies the no-signalling condition~\eqref{eq:nosig}
\be
\label{eq:mar}\forall a,b,x,y \quad \Prob[a|x,y] = \sum\limits_{b} \Prob[a,b|x,y] = \frac{1}{2}
\ee
As a result, it is impossible for them to use the PR-box to communicate instantly. 
Yet, if at a later time Alice and Bob meet or exchange information about the outcomes they have got, they will find that $a,b,x$ and $y$ always satisfy, equation \eqref{eq:prnice}.

\section{No-signalling correlations}

In a no-signalling world, we would a priori expect that indeed all no-signalling correlations are realizable. That is, there exists a box similar to the PR-box 
to which we can give inputs $x$ and $y$, and it generates some outputs $a$ and $b$ according to a desired no-signalling distribution $\Pr[a,b|x,y]$. 
Here, we show that as long as there exist uncomputable functions (see below) in such a theory, not all distributions can be realized, even if they obey the 
no-signalling principle. That is, the set of allowed distributions is strictly smaller than dictated by the no-signalling principle alone.

In order to show this, we will now explain a procedure which demonstrates how any function in principle can be computed in a distributed fashion, i.e., by a PR-type box.
In~\cite{linden:advantage} Linden et. al. describe a form of distributed non-local computation. However in their method Alice and Bob themselves do not know the actual inputs, rather they are given an additive sharing of the inputs~\cite{linden:advantage,brunner:distillation}.
To get some intuition of how our construction works in contrast, let us first consider the PR-box in a more operational way. In particular, we will think about a procedure
according to which the box generates the outputs. Note that the box does of course not follow an internal procedure, but is merely given by a set of probability distributions.
Yet, this view provides us with an operational perspective on how to write down such distributions. More precisely, we imagine that
\begin{enumerate}
\item Upon receiving the inputs $x$ and $y$ from Alice and Bob, the box select an independent and uniformly random bit $r\in_R \{0,1\}$.
\item Then, it computes $x\cdot y$.
\item Finally, it assigns $a\gets r$ and $b\gets r\oplus x\cdot y$.
\end{enumerate}
To see why this procedure works, first observe that
\be
a \oplus b = r \oplus (r\oplus x\cdot y) = x\cdot y.
\ee
And, as $r (=a)$, is an independently generated random bit, $a$ must not be affected by the input $y$ from Bob. On the other side, Bob receives $b = r \oplus x\cdot y$. To him, it is also completely random because it is encrypted with a random bit $r$ which is generated independently. So, he also does not have any information about the input $x$ from Alice. As a result, 
the box is no-signalling. 

\begin{figure}
\centering
\begin{pspicture}(1.1,0.5)(7,5)
	\psframe[linewidth=2pt](1.8,1.8)(6.4,3.2)
	\rput(1.2,2.50){Alice}
	\rput(6.9,2.50){Bob}
	\psline[linewidth=2pt,linearc=1]{->}(1.2,4)(1.8,3.8)(2.2,3.2)
	\rput(1,4){$x$}
	
	\psline[linewidth=2pt,linearc=1]{->}(7,4)(6.4,3.8)(6,3.2)
	\rput(7.2,4){$y$}
	\rput(4.1,2.80){$r\in_R\{0,1\}$}
	\rput(2.5,2.10){$a\gets r$}
	\rput(5,2.10) {$b\gets r \oplus f(x,y)$}
	\psline[linewidth=2pt,linearc=1]{->}(2.2,1.8)(2.1,1.5)(1.8,1.2)
	\rput(1.6,1){$a$}
	\psline[linewidth=2pt,linearc=1]{->}(6,1.8)(6.1,1.5)(6.4,1.2)
	\rput(6.6,1){$b$}
\end{pspicture}
\caption{No-signalling computation of the function $f(x,y)$ }\label{fig:box}
\end{figure}
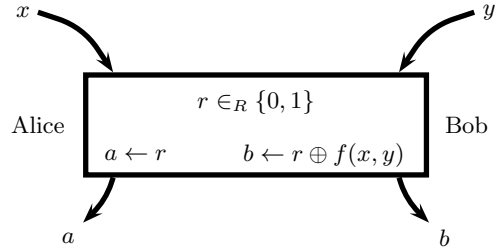

We have described how a PR-Box can operate in principle. From this we can have a generalized version of it. Namely, we can have a box that satisfies:
\be
\label{eq:prgen}a \oplus b = f(x, y),
\ee
where $f:\{0,1\}\times\{0,1\} \to \{0,1\}$ is any function of $x$ and $y$. We achieve this by simply, computing $f(x,y)$ in step $2$ and assigning $a\gets r$ and $b\gets r\oplus f(x,y)$ in step $3$. We call this the \emph{no-signalling computation} of function $f(x,y)$, because by construction this computation is also no-signalling. In fact, by the same argument we can generalize this box farther and non-locally compute any function $f:\{0,1\}^l\times\{0,1\}^m \to \{0,1\}$, which takes strings $x$ and $y$ of arbitrary lengths as inputs from Alice and Bob accordingly and computes  $f(x,y)$ non-locally still satisfying the non-signaling condition. In fact, W. van Dam shows in \cite{wim:nonlocal} that one can also perform this generalized no-signalling computation of $f(x,y)$ by using multiple copies of 
PR-Boxes~\footnote{Note that our result does \emph{not} exclude the existence of PR-boxes. When it comes to no-signalling computations, PR-boxes can be understood as the analog of an AND-gate in a computational circuit - the fact that some functions cannot be computed does not exclude the possibility of such elementary gates.}.

As we have seen, if any distribution is allowed which is compatible with the no-signalling principle, then \emph{any} function can be computed in such a non-locally distributed manner. That is, once Alice and Bob have their outcomes $a$ and $b$ respectively, they can meet or exchange this values at any later time and recover the functions value by computing $a \oplus b = f(x,y)$. This means that \emph{if} the function $f$ is not computable in a world where PR-boxes exist, then also such correlations cannot exist - even if they obey the principle of no-signalling.

\section{Computability}

Let us now discuss in more detail what it means for a function to be computable.
Intuitively, a function is computable in some particular physical theory, if this theory allows us to build a machine that takes as inputs $x$ and $y$, and 
outputs $f(x,y)$ after a finite amount of time.
Classically, it turns out that according to the \emph{Church-Turing thesis} \cite{church:thesis,turing:thesis}, 

\emph{Every `function which would naturally be regarded as computable' can be computed by the universal Turing machine.} 

In \cite{deutsch:uqc} Deutsch states a stronger version of this thesis, known as the Church-Turing \emph{principle}:

\emph{Every finitely realizable physical system can be perfectly simulated by a universal model computing machine operating by finite means.}

Intuitively, a universal model computing machine is a machine that can simulate any other computing machine. 
To understand this concept, let us consider the example of the 
Quantum turing machines~\cite{deutsch:uqc}. Let us call this machine $U_q$. We call this machine \emph{universal}, because it can be programmed to simulate all the other physically realizable computing machines operating under quantum mechanics, including itself. That is, for any machine $M$ computing a function $f$ there is a program $P$ for the universal quantum turing machine $U_q$ which can also compute the function. In fact, it has been shown that even a \emph{classical} turing machine~\cite{turing:thesis} can simulate the universal quantum turing machine with an additional overhead~\cite{deutsch:uqc}. One can thus consider the program $P$ to be a string on a suitable finite alphabet. 

It is shown that not all functions are computable by turing machines, and we briefly sketch the argument here~\cite{LP:textbook}:
Without loss of generality we can assume $P \in \01^n$ and the decision function to be $f:\01^* \to \01$. That is, the program is a finite string of $0$s and $1$s of arbitrary length $n$. This allows us to find a one-to-one correspondence between all the possible programs to the set of natural numbers $N$ by ordering the program strings in lexicographic order. That is the set of possible programs is countably infinite. On the other hand, the set of possible functions $f:\01^* \to \01$ is not countable. As a result no matter how powerful a computing machine is, as long as it is programmed using finite strings there will always remain functions not computable by that computing model. And by Church-Turing Principle such computations cannot be physically realized in quantum mechanics. 

It is important to note that finite in the statement of the Church-Turing principle means that for any particular computation we only need a finite amount of computational 
resources~\footnote{A machine that looks up the answer in a list would need an infinite list to produce the answer for arbitrary inputs.}. 
In principle, an unlimited amount of resources
\emph{are} available, however only a finite amount are used at any given time (see~\cite{deutsch:uqc} for details). 
Further details on physical uncomputability can be found in e.g.~\cite{deutsch:uqc, nielsen:measurements, nielsen&chuang:qc}. A treatment of classical uncomputability can be found in e.g. the following textbook~\cite{LP:textbook}.

Classically and quantumly, 
one uncomputable function is Turing's halting function $H(x,y)$ \cite{turing:thesis}. This function takes as inputs $x$ which is the binary encoding of any program and $y$ which is the input to the program $x$ and computes a boolean output indicating whether program $x$ halts on input $y$ or keeps running forever. 
It was shown by Alan Turing that one cannot compute this function $H$. That is, no matter how much resource (time and memory) is provided, if someone assumes the solution of this function one will immediately lead to a contradiction.
Now, if in our construction we take $f = H$, we see that Alice and Bob upon communicating one bit in the end, can compute the Halting function non-locally. Note that for this construction to work at least one of the input sets  $\setX$ and $\setY$ has to be infinite. ~\footnote{This construction works even if one of the input sets $\setX$ and $\setY$ is fixed and finite. As long as at least one of them is countably infinite, we can consider that as the set of programs and the other as inputs. It is known whether a program halts even on a fixed input is undecidable~\cite[Theorem 5.4.2]{LP:textbook}.}.

In a general theory, the input, i.e., program, to a universal computing machine may of course not only consist of classical bits, but also of more 
generalized bits allowed in that particular theory. However, for the special 
case of the halting function, Turing's argument showing that it is impossible to compute the Halting function only hinges on the fact that the machine is indeed universal and the program has a finite description \emph{within} that theory. Hence, if the Church-Turing principle would hold in that theory, the halting function should also be uncomputable.

Of course a world, where any non-signaling correlation is allowed, has many unusual and unexpected properties~\cite{versteeg:uncertainty,pawlowski:infoCausality,b3ew:nosignaling,js:uncertainty}. So it is not inconceivable that also the Church-Turing principle should break down. 
It could be, for example, that there does \emph{not} exist a universal computing machine that can simulate any other machine in that physical theory, or in other words,
not all finitely realizable processes are simulatable. We emphasize that we make no statement about whether the Church-Turing principle holds in any other theories. Our result merely implies that we must make a choice: either the Church-Turing principle \emph{does} hold and some functions are not computable, but then it imposes an additional limit on possible non-local correlations. Or, all correlations compatible with the no-signalling principle are allowed, but then all functions~\footnote{Note that one can always split argument to make a multi-valued function.} are computable and the Church-Turing principle does not hold~\footnote{
We also emphasize that whenever the Church-Turing holds, one cannot even write down the no-signalling distribution corresponding to the the box in Figure~\ref{fig:box} for all possible inputs $x$ and $y$ as this would be equivalent to computing the function manually for all inputs. Note, however, that just with the halting function itself we do not need to write down the function in order to achieve a contradiction. Rather, one argues that \emph{if} the halting function were computable, then we obtain a contradiction.}.

\section {Approximate Cases}

Let us now consider the case where 
our correlations are non-signalling, but not \emph{super strong}. That is, we consider approximate versions of PR-type boxes for which for some probability $p>\frac{1}{2}$ we have,
\be
\label{eq:weak} \forall a,b, x,y\quad p\leq\Prob[a\oplus b=f(x,y)] < 1\ .
\ee
As a result by computing $a \oplus b$ Alice and Bob can compute the function $f$ with probability at least $p$ of giving the correct outcome. Let's denote this by $f_p(x,y)$.
Quantum mechanics can provide us with such approximate boxes.
For example, we know that for $x,y \in \01$ the function $f(x,y) = x \cdot y$ can be computed in a distributed manner with $p = 1/2 + 1/(2\sqrt{2}) \approx 0.85$ for all $x$ and $y$
-- this is just the quantum violation of the CHSH inequality. For any $p$, we can thus again ask the question: should we expect such approximate boxes to exist?

Evidently, for any $p > 1/2$ it is possible to approximate any computable function with correctness $1-\epsilon$ for any $\epsilon >0$ by repeating the process many times. 
Yet, one can also show directly that there exist undecidable problems for which even such probabilistic versions cannot be computed. That is, one cannot compute that function with some constant probability $p$ of correctness. 
As an example, let us consider the probabilistic halting function $H_p(x,y)$ which tells us with probability of correctness $p$ whether program $x$ halts on input $y$ with 
probability $p$. Using a similar argument that was used by Turing to prove the uncomputability of the deterministic halting function, it can be shown that even this probabilistic function cannot be computed~\cite[Exercise 3.6]{nielsen&chuang:qc}.

We thus conclude that the question of computability imposes constraints on non-local correlations even for the approximate case of $1 > p > 1/2$~\footnote{Note that $p$ here is not an average as in the CHSH inequality~\eqref{eq:chsh}. That is, here we require for each pair of $x$ and $y$, the probability $p$ of getting outcome $a\oplus b = f(x,y)$ to be strictly greater than $1/2$.}.

\section{Discussion}

We have shown that \emph{if} the Church-Turing principle holds in a particular physical theory, then it imposes additional constraints on the existence of non-local correlations. Here we emphasize that this result holds for any general theory that is constrained by no-signalling principle. 

For example, in a world where PR-type boxes~\cite{popescu:nonlocal,popescu:nonlocal2,popescu:nonlocal3} exist, we previously worked with the assumption that all non-local correlations 
are allowed, as long as they obey the no-signalling principle.
However, we now know that there \emph{may} be further constraints:  either the Church-Turing principle does not hold, or non-local correlations really are strictly more limited -- at least if Alice and Bob can meet again to combine their inputs. 
Of course, locally our box from Figure~\ref{fig:box} yields fully random outputs and neither Alice nor Bob can determine $f(x,y)$ on their own. 
This means that locally their outcomes are not constrained by the question whether $f$ is computable.

In the quantum case, our result should be compared to~\cite{nielsen:measurements} in which it is shown that if the Church-Turing principle holds, then there are measurements and unitary operators that cannot be performed on a local quantum system. Here, we show that indeed in \emph{any} theory in which the Church-Turing principle holds, certain states and/or measurements
are not available to us as otherwise any (approximate) no-signalling computation could be performed. As such, our result can also be understood as imposing a limit on the kind of states and measurements permissible. 

It is an interesting open question to identify the class of theories for which the Church-Turing principle holds, and thus imposes a strict limit on physical observations.

\acknowledgments

{\bf Acknowledgments} We thank CQT's nonlocal club for interesting discussions and comments.
This research was supported by the National Research Foundation and Ministry of Education, Singapore.

\end{document}